\documentstyle[12pt,epsfig]{article}
\def\Lc#1{{\rm Lc}_{#1}}

\def\Ld#1{{\rm Ld}_{#1}}

\def\Li2{{\rm Li_2}}
\def\A{{\cal A}}
\def\L{{\cal L}}
\def\Lhat{\hat {\cal L}}
\def\T{{\cal T}}
\def\M{{\cal M}}
\def\P{{\cal P}}
\def\e{\epsilon}
\begin{document}
\begin{titlepage}
\begin{flushright}
DTP/97/44\\
RAL TR 97-027\\
hep-ph/9706297\\
\end{flushright}
\vspace{1cm}
\begin{center}
{\Large\bf The One-loop QCD Corrections for $ \gamma^* \to q\bar q gg $}\\
\vspace{1cm}
{\large
J.~M.~Campbell, E.~W.~N.~ Glover}\\
\vspace{0.5cm}
{\it
Physics Department, University of Durham,\\
Durham DH1~3LE,
England} \\
\vspace{0.5cm}
and \\
\vspace{0.5cm}
{\large D.~J.~Miller}\\
\vspace{0.5cm}
{\it
Rutherford Appleton Laboratory,  Chilton, \\
Didcot, Oxon,
OX11 0QX, England}

\vspace{0.5cm}
{\large June 1997}
\vspace{0.5cm}
\end{center}
\begin{abstract}
We compute the
one-loop QCD amplitudes for the decay of an off-shell
vector boson with vector couplings into a quark-antiquark pair 
accompanied by two gluons
keeping, for the first time, all orders in the number of colours.
Together with previous work this completes 
the calculation of the necessary one-loop amplitudes needed
for the calculation of the
next-to-leading order ${\cal O}(\alpha_s^3)$ corrections to
four jet production in electron-positron annihilation,
the production of a gauge boson accompanied by two jets in
hadron-hadron collisions and three jet production in deep inelastic
scattering. 
\end{abstract}
\end{titlepage}

Multi-jet events in electron-positron annihilation have 
long been a source of vital information about the structure of QCD.
In 1979, three jet events at DESY gave the first clear indications 
of the existence of the gluon \cite{gluon}, while more recently, 
the colour factors that determine the gauge group of QCD have been 
determined \cite{casimir}.
While the general features of events have been well described by 
leading order 
predictions, many quantities are known to suffer large radiative 
corrections.
For example, the experimental four jet rate measured as a function 
of the jet resolution parameter $y_{\rm cut}$, was significantly 
higher than the leading order expectation.   
This is of particular interest at LEP II energies where 4 jet events
are the main background to $WW$ events where both $W$ bosons decay 
hadronically.

Recently there has been much progress towards a more complete 
theoretical description of four jet events produced in 
electron-positron annihilation.
Dixon and Signer \cite{DS} have created a general purpose Monte Carlo program 
to evaluate the next-to-leading order corrections to a wide variety of 
four-jet and related observables.
Such numerical programs have three important ingredients; the one-loop 
four parton amplitudes, the tree-level five parton amplitudes and a 
method for combining the four and five parton final states together.
The tree-level amplitudes for $e^+e^- \to q\bar q ggg$ and 
$e^+e^-\to q\bar q Q\bar Q g$  are well known \cite{HZ,BGK,FGK} 
and several 
general methods have been developed for numerically isolating 
the infrared divergences \cite{slice,subtract,hybrid}.
However, the one-loop amplitudes are not fully known, and the 
next-to-leading order corrections described by the Dixon-Signer program 
\cite{DS} are therefore incomplete.

The one-loop amplitudes relevant for this process are
$e^+e^- \to q\bar q gg$ and $e^+e^- \to q\bar q Q\bar Q$.
Two groups \cite{GM,BDKW} have computed the amplitudes for the 
four quark final states  keeping all orders in the number of 
colours, and these matrix elements have been implemented in 
the Dixon-Signer Monte Carlo \cite{DS}.
The situation for the two quark two gluon process is less 
complete and may be illustrated by examining the colour 
structure  of the one-loop contribution to the four-jet cross section,
\begin{equation}
\sigma_{{\rm q\bar q gg}}^{{\rm 1-loop}} = N^2(N^2-1) \left(
 \sigma^{(a)} 
+ \frac{1}{N^2} \sigma^{(b)} 
+ \frac{1}{N^4} \sigma^{(c)} \right),
\end{equation}
where $N$ is the number of colours.
Bern, Dixon and Kosower \cite{BDKZ2q2g} have provided 
compact analytic results for the leading colour part of the two quark 
two gluon process ($\sigma^{(a)}$) 
which should account for approximately 90\% of the next-to-leading 
order corrections.  Indeed,  keeping all of the known contributions, 
Dixon and Signer find that the next-to-leading order predictions 
match onto the experimental data much better \cite{DS}.  However, 
the subleading terms  $\sigma^{(b)}$ 
and $\sigma^{(c)}$ are still expected to contribute a further 
10\% to the 4-jet rate.
In this letter, we compute the one-loop matrix elements for the 
$\gamma^* \to q\bar q gg$ process keeping,
for the first time,
all orders in the number of colours.
These results can be implemented directly into the Dixon-Signer 
Monte Carlo \cite{DS}, or can be used as a cross check with the results 
anticipated from the helicity approach \cite{BDKforthcoming}.
 
As in the calculation of the one-loop four quark matrix elements \cite{GM} 
and in contrast to the helicity method employed by Bern et al. \cite{BDKW},
we compute `squared' matrix elements, i.e. the
interference between tree-level and one-loop amplitudes. 
By doing so, we reduce all tensor integrals down to scalars directly
leaving at worst three (tensor) loop momenta in the
numerator of the box integrals.

The evaluation of the tensor integrals represents the bulk of the
calculation and should be treated with some care. The `squared'
tree-level matrix element for a given process contains inverse
powers of certain invariants, resulting from the propagators present.
This singularity structure is largely responsible for the infrared behaviour
of the matrix elements. 
Based on general arguments, one expects that 
the one-loop corrections should have 
the same singularity structure as tree-level and should not  
contain more inverse powers of these (or other) invariants
or additional kinematic singularities resulting from the
calculational scheme used. 
However, as well as generating extra inverse powers of the 
invariants (above and beyond the expected tree-level singularities), 
traditional methods of reducing the 
tensor integrals to combinations of scalar integrals introduce (many)  
factors of the Gram determinant ($\Delta$) in the denominator.
Although the matrix elements themselves are finite as $\Delta \to 0$,
individual terms appear to diverge.
These apparent singularities can be
avoided by forming suitable combinations of the scalar integrals
which are finite in the limit of vanishing
$\Delta$ as well as in the limit where the invariant mass 
of a pair of particles tends to zero. Motivated by the work of Bern et al,
\cite{BDK1},  
we have shown  in ref.~\cite{CGM} how these 
functions are naturally obtained by differentiation
of the original scalar loop-integral.
The
advantages of this decomposition are then two-fold. First, the expressions
for the one-loop matrix elements are significantly simplified by grouping integrals together.
Second,
we can ensure that 
the one-loop singularity structure reproduces that at tree level.
In other words, the only allowed kinematic
singularities are 
those appearing at tree level, multiplied by coefficients 
that are well behaved in all kinematic limits. This latter 
point is achieved by using identities amongst the set of 
basic functions and tends not to reduce the size of the 
answer, although  
the absence of further kinematic
singularities ensures that the matrix elements are numerically stable.
There is a slight price to pay in that, compared to the raw 
scalar integrals, the
the set of basic functions is now enlarged and is not linearly
independent, resulting in some ambiguity in the form of the final
result.
To generate and simplify our results,
we have made repeated use of the algebraic manipulation packages
FORM and MAPLE.

For the process under consideration, $\gamma^* \to q\bar q gg $ 
we label the momenta as,
\begin{equation}
\gamma^*(p_{1234}) \to q(p_1) + \bar q(p_2) + g(p_3) + g(p_4),
\end{equation}
and systematically eliminate the
photon momentum in favour of the four massless parton momenta.

The colour structure of the matrix element at tree-level $(n=0)$ and
one-loop $(n=1)$ is rather simple and we have,
\begin{eqnarray}
\lefteqn{\M^{(n)} =  \epsilon^\mu
\epsilon^{\nu_3} 
\epsilon^{\nu_4}  
\M_{\mu\nu_3\nu_4}^{(n)} = e g_s^2
\left(\frac{g_s}{4\pi}\right)^{2n}}\nonumber
\\
&\times & \left\{
 (T^{a_3} T^{a_4})_{c_1c_2} \A^{(n)}_1(3,4)
 + (T^{a_4} T^{a_3})_{c_1c_2} \A^{(n)}_1(4,3)
   + \frac{1}{2} \delta_{c_1c_2}\delta_{a_3a_4} \A^{(n)}_2(3,4)  \right\},
\end{eqnarray}
where $c_1$, $c_2$ are the colours of the quarks and $a_3$, $a_4$ the colours
of the gluons.
The arguments of $\A_i$ indicate a permutation of the momenta of the
external gluons. At lowest order,
\begin{equation}
\A^{(0)}_2(3,4)=0,
\end{equation}
while at one-loop we find,
\begin{eqnarray}
\A^{(1)}_1(3,4) &=& N \A^{(1)}_A(3,4)
 - \frac{1}{N} \A^{(1)}_B(3,4) , \\
\A^{(1)}_2(3,4) & \equiv & \A^{(1)}_C(3,4) \equiv \A^{(1)}_C(4,3).
\end{eqnarray}
The functions $A^{(1)}_\alpha(i,j)$, $\alpha = A,B,C$ represent the
contributions of the three gauge invariant sets of Feynman diagrams
shown in Figs.~1,~2 and 3 respectively.
We note that the contribution $\A^{(1)}_A(3,4)$ also includes terms
proportional to $n_F/N$, where $n_F$ is the number of light quark flavours.
These terms originate from the fermionic self-energy and triple gluon vertex
corrections shown in the last two diagrams of Fig. 1.
The Dixon-Signer Monte Carlo \cite{DS} incorporates the leading colour term, 
$\A^{(1)}_A(3,4)$.

\begin{figure}[thbp]
\begin{center}
~\epsfig{file=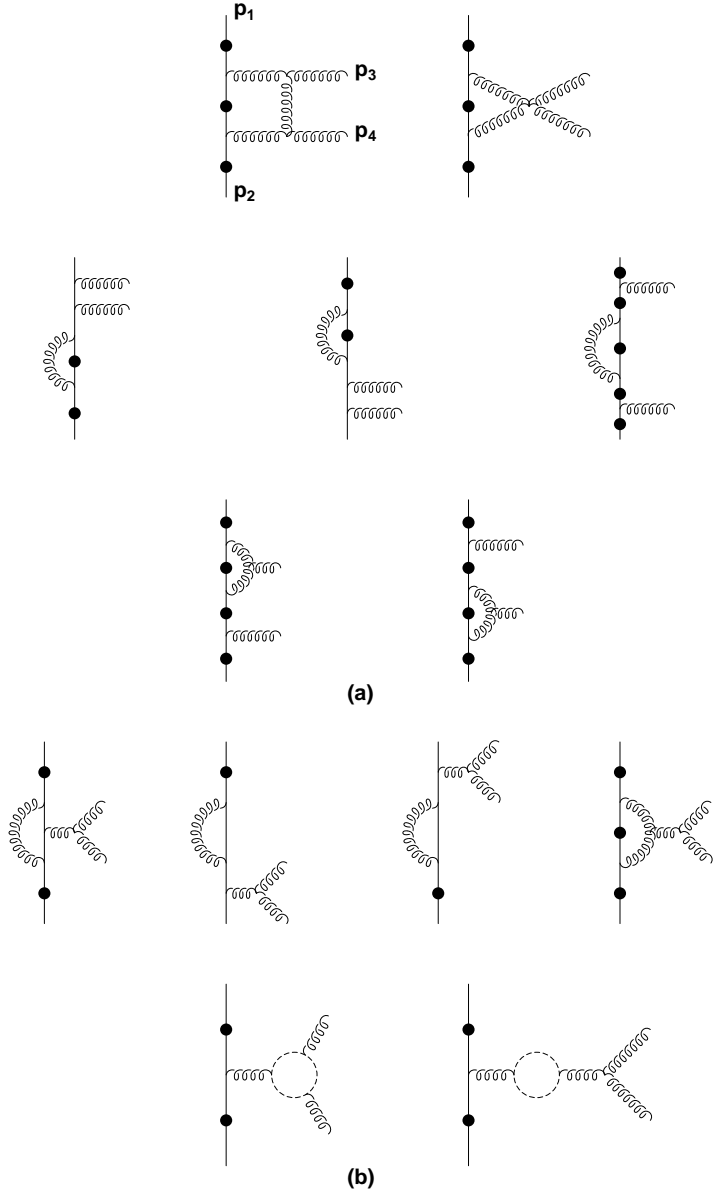,height=16cm}
\end{center}
\caption[]{The classes of Feynman diagrams relevant for the function
$\A^{(1)}_A(3,4)$.
Reading clockwise round the diagram and starting from the quark ($p_1$) at the top,
we encounter gluon ($p_3$)  before gluon ($p_4$) and end at antiquark ($p_2$). 
The solid circle indicates the possible positions for 
attaching the off-shell photon
to the quark-antiquark pair. 
Diagrams (a), taken with both permutations of gluons 3 and 4, contribute
to the piece $\L_A$ while the permutation shown in 
(a)+(b) gives the contribution to $\L_A(3,4)$. 
Diagrams with self-energy corrections on the external lines are zero in 
dimensional regularisation and have been omitted. }
\end{figure}

\begin{figure}[thbp]
\begin{center}
~\epsfig{file=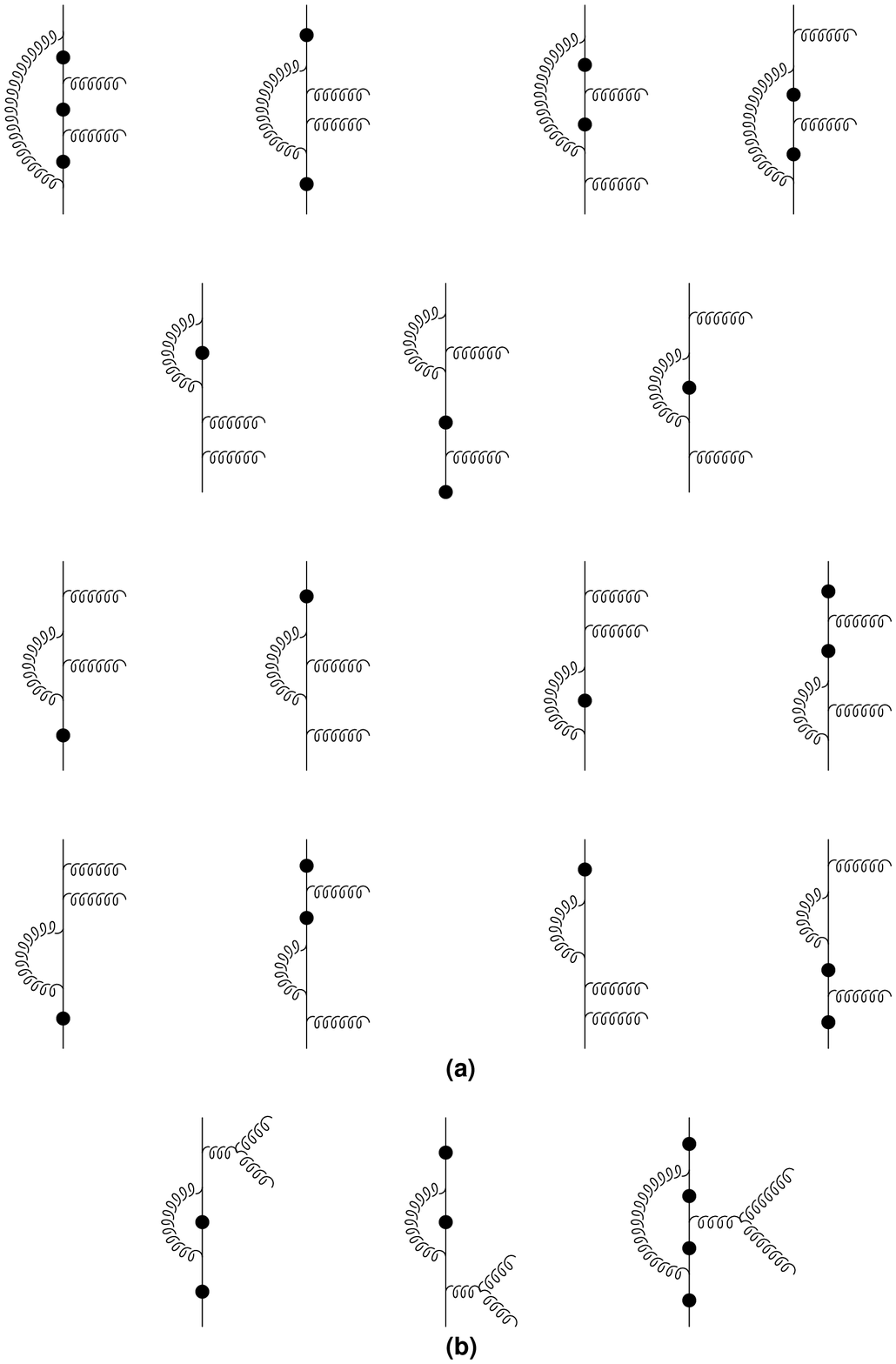,height=16cm}
\end{center}
\caption[]{The classes of Feynman diagrams relevant for the function
$\A^{(1)}_B(3,4)$ 
Reading clockwise round the diagram and starting from the quark ($p_1$) at the top,
we encounter gluon ($p_3$)  before gluon ($p_4$) and end at antiquark ($p_2$). 
The solid circle indicates the possible positions for 
attaching the off-shell photon
to the quark-antiquark pair. 
Diagrams (a), taken with both permutations of gluons 3 and 4, contribute
to the piece $\L_B$ while the permutation shown in 
(a)+(b) gives the contribution to $\L_B(3,4)$. }
\end{figure}

\begin{figure}[th]
\begin{center}
~\epsfig{file=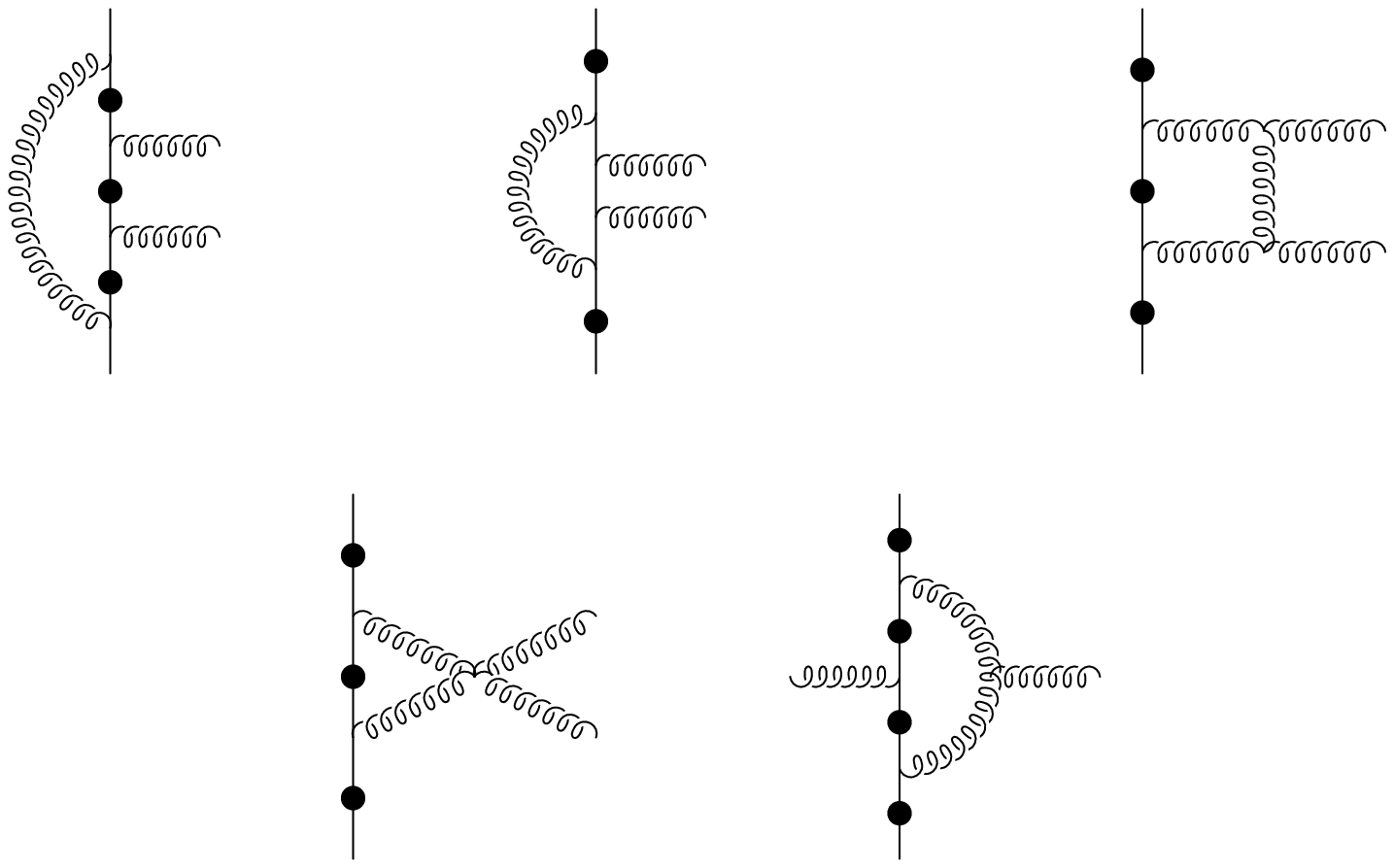,height=7cm}
\end{center}
\caption[]{The classes of Feynman diagrams relevant for the function
$\A^{(1)}_C$ when taken with both permutations of the gluons.
Reading clockwise round the diagram and starting from the quark ($p_1$) at the top,
we encounter gluon ($p_3$)  before gluon ($p_4$) and end at antiquark ($p_2$). 
The solid circle indicates the possible positions for 
attaching the off-shell photon
to the quark-antiquark pair. }
\end{figure}

The squared matrix elements are relatively straightforwardly obtained
at leading order \cite{ERT}. Some care must be taken to ensure that
only physical, transverse degrees of freedom are included for the
external gluons. We do this by requiring that terms in 
$\M_{\mu\nu_3\nu_4}^{(0)}$ that vanish when contracted with the
physical gluon polarisation vectors $\epsilon^{\nu_3}$ or
$\epsilon^{\nu_4}$ are removed by hand \cite{combridge}. 
The sum over the polarisations of all the gauge bosons may then be
performed in the Feynman gauge,
\begin{equation}
\sum_{\rm spins} \epsilon^\mu \epsilon^{*\nu} = -g^{\mu\nu}.
\end{equation}
Hence,
\begin{equation}
\sum_{\rm spins} |\M^{(0)}|^2  = \frac{e^2 g_s^4}{4}(N^2-1)N
\Biggl \{
\T(3,4) + \T(4,3) - \frac{1}{N^2} \T \Biggr \},
\end{equation}
where,
\begin{equation}
\T(3,4) = \sum_{\rm spins}
\Big| A_1^{(0)\dagger}(3,4) \A_1^{(0)}(3,4) \Big|,
\end{equation}
and,
\begin{equation}
\T = \sum_{\rm spins}
\Big | \left( A_1^{(0)\dagger}(3,4)+A_1^{(0)\dagger}(4,3) \right)
  \left( \A_1^{(0)}(3,4)+\A_1^{(0)}(4,3) \right) \Big|.
\end{equation}
The 3-gluon vertex contributions to $\A_1^{(0)}(3,4)$ and $\A_1^{(0)}(4,3)$
enter with opposite sign, so $\T$ (with no arguments) is the contribution
from the pure QED-like diagrams.

The relevant `squared' matrix elements are the interference between
the tree-level and one-loop amplitudes,
\begin{eqnarray}
\lefteqn{
\sum_{\rm spins} 2 |\M^{(0)\dagger}\M^{(1)}|  = \frac{e^2g_s^4}{4}
\left(\frac{\alpha_sN}{2\pi}\right)(N^2-1)N}\nonumber
\\
&\times& \Biggl \{
\L_A(3,4) + \L_A(4,3)-\frac{1}{N^2} \biggl( \L_A+\L_B(3,4)+\L_B(4,3)
-\L_C \biggr)+\frac{1}{N^4} \L_B \Biggr \},
\end{eqnarray}
with,
\begin{equation}
\L_\alpha(3,4) = \sum_{\rm spins}
\Big | A^{(1)\dagger}_\alpha(3,4) \A_1^{(0)}(3,4) \Big |,
\end{equation}
for $\alpha = A,~B$ and the QED-like structures,
\begin{equation}
\L_\alpha = \sum_{\rm spins}
\Big | \left( A^{(1)\dagger}_\alpha(3,4)+A^{(1)\dagger}_\alpha(4,3) \right)
\left( \A_1^{(0)}(3,4)+\A_1^{(0)}(4,3) \right) \Big |.
\end{equation}
Hence the `squared' matrix elements are described by 5 independent
$\L_\alpha$.  The $\L_\alpha(3,4)$ are symmetric under 
the exchange $p_1 \leftrightarrow p_2$ 
{\bf and} $p_3 \leftrightarrow p_4$
while the $\L_\alpha$ are symmetric under either
$p_1 \leftrightarrow p_2$ {\bf or} $p_3 \leftrightarrow p_4$.

Working in dimensional regularisation with $4-2\e$ spacetime dimensions, 
it is straightforward to remove the infrared and ultraviolet
poles from the $\L_\alpha$ since they are proportional to the tree-level 
amplitudes,
\begin{eqnarray}
\L_A(3,4) &=&  
  \left( -\frac{\P_{13}}{\e^2} - \frac{\P_{24}}{\e^2}
- \frac{\P_{34}}{\e^2} - \frac{3}{2} \frac{\P_{1234}}{\e}
 \right) \T(3,4) + \Lhat_A(3,4),  \\
\L_A &=&  
  + \frac{1}{2}\left( -\frac{\P_{13}}{\e^2} - \frac{\P_{24}}{\e^2}
 - \frac{\P_{34}}{\e^2} - \frac{3}{2} \frac{\P_{1234}}{\e}\right)
  \left(\T + \T(3,4)-\T(4,3)\right) \nonumber
\\
 && + \frac{1}{2}\left( -\frac{\P_{14}}{\e^2} - \frac{\P_{23}}{\e^2}
 - \frac{\P_{34}}{\e^2} - \frac{3}{2} \frac{\P_{1234}}{\e}\right) 
  \left(\T + \T(4,3)-\T(3,4)\right)
 + \Lhat_A,  \\
\L_B(3,4) &=&  
  \left( -\frac{\P_{12}}{\e^2} - \frac{3}{2} \frac{\P_{1234}}{\e}
 \right) \T(3,4) + \Lhat_B(3,4),  \\
\L_B &=&  
  \left( -\frac{\P_{12}}{\e^2} - \frac{3}{2} \frac{\P_{1234}}{\e}
 \right) \T + \Lhat_B, \\
\L_C &=&  
  + \frac{1}{2}\left( -\frac{\P_{12}}{\e^2} - \frac{\P_{34}}{\e^2}
 + \frac{\P_{14}}{\e^2} + \frac{\P_{23}}{\e^2} \right)
 \left(\T + \T(3,4)-\T(4,3)\right) \nonumber \\ 
 &&+ \frac{1}{2}\left( -\frac{\P_{12}}{\e^2} - \frac{\P_{34}}{\e^2}
 + \frac{\P_{13}}{\e^2} + \frac{\P_{24}}{\e^2} \right)
 \left(\T + \T(4,3)-\T(3,4)\right) 
 + \Lhat_C.
\end{eqnarray}
where we have introduced the notation,
\begin{equation}
\P_{x} = \left(\frac{4\pi\mu^2}{-s_{x}}\right)^{\e}
\frac{\Gamma^2(1-\e)\Gamma(1+\e)}{\Gamma(1-2\e)}.
\end{equation}
In physical cross sections, this pole structure cancels with
the infrared poles from the $\gamma^* \to q\bar qgg +g$ 
process and those generated by ultraviolet renormalisation.

In determining the finite pieces, $\Lhat$,
we are concerned to ensure that the singularity structure matches that 
of the tree-level functions $\T(3,4)$ and $\T$.
In terms of the generalised
Mandelstam invariants,
\begin{equation}
s_{ij}=(p_i+p_j)^2, \qquad s_{ijk}=(p_i+p_j+p_k)^2,
\qquad s_{ijkl}=(p_i+p_j+p_k+p_l)^2,
\end{equation}
$\T$ contains single poles in $s_{13}$, $s_{23}$, $s_{14}$ and $s_{24}$
while $\T(3,4)$  has poles in $s_{13}$, $s_{34}$ and $s_{24}$.
In addition, both $\T$ and $\T(3,4)$
contain double poles in the triple invariants $s_{134}$ and $s_{234}$.
Using the tensor reduction described in \cite{CGM}, possible
singularities due to Gram determinants are automatically protected.
However, it is possible to generate apparent singularities in 
double or triple invariants such as $s_{12}$ or $s_{123}$.
These poles do not correspond to any of the allowed 
infrared singularities and
the matrix elements are finite as, for example, 
$s_{12} \to 0$ or $s_{123} \to 0$.
In fact, it is straightforward to explicitly remove such poles using 
identities amongst the combinations of scalar integrals.
For example,  the identity,
\begin{equation}
\frac{(s_{123}-s_{12})}{s_{123}} \Ld{22}(p_{12},p_3,p_4)
 = - \Ld{21}(p_{12},p_3,p_4)
   + \Ld{1}(p_{12},p_3,p_4)
   + \Lc{1}(p_{12},p_{34}),
\end{equation}
relating the functions for box integral functions with 
two adjacent masses (defined in Appendix~B of
\cite{CGM}) is useful to eliminate poles in $s_{123}$.
The finite pieces can be written symbolically as,
\begin{equation}
\Lhat=\sum_i P_i(s) {\rm L}_i,
\end{equation}
where the coefficients $P_i(s)$ are rational polynomials of 
invariants.
The finite functions ${\rm L}_i$ are the linear combinations of
scalar integrals defined in \cite{CGM} 
which are well-behaved in all kinematic limits.
Any denominators of the corresponding tree-level matrix element
are allowed in the coefficient $P_i(s)$, with any additional fake
singularities protected by ${\rm L}_i$. Typically the coefficient
of a given function contains ${\cal O}(100)$ terms, comparable
with the size of the tree-level matrix elements. The number of
functions is rather large, of ${\cal O}(100)$, but is a minimal
set which protects all the kinematic limits and is therefore
numerically stable.
The analytic expressions for the
individual $\Lhat_\alpha$, and particularly the subleading colour terms 
$\Lhat_B$ and $\Lhat_C$, are therefore rather lengthy so 
we have constructed a FORTRAN subroutine to evaluate them 
for a given phase space point.
This can be directly implemented in general purpose next-to-leading order
Monte Carlo programs for the
$e^+e^- \to 4~{\rm jet}$,
$e^\pm p \to e^\pm + 3~{\rm jet}$ and
$p\bar p \to V + 2~{\rm jet}$  processes.

To summarize,
we have performed the first calculation of the one-loop `squared'
matrix elements for the $\gamma^* \to q\bar q gg$ process
keeping all orders in the number of colours.
We have grouped the Feynman diagrams according to the
colour structure, but unlike the helicity approach of \cite{BDK3},
have used conventional dimensional regularisation throughout.
We have used the one-loop reduction method described in \cite{CGM}
to obtain the finite parts of the one-loop matrix amplitudes
in terms of functions that are well behaved in all kinematic limits.
The finite one-loop expressions $\Lhat_\alpha$ contain the same 
singularity structure as the tree level amplitudes $\T$ and are 
numerically stable.
Our results are rather lengthy and we have provided a FORTRAN 
implementation of them that can be either used with the existing 
Dixon-Signer program for
$e^+e^- \to 4$~jets or as a completely independent check of amplitudes 
obtained 
using the helicity approach \cite{BDKW,BDKZ2q2g,BDKforthcoming}.
Together with previous work \cite{GM,BDKW,BDKZ2q2g} this completes 
the calculation of the necessary one-loop amplitudes for the coupling of an
electroweak gauge boson to four massless partons with the following
 exceptions:
\vspace{-0.5cm}
\begin{itemize}
\item[{(a)}] Contributions proportional to the axial coupling. 
\item[{(b)}] Contributions where the electroweak boson couples to a closed fermion loop.
\end{itemize}
\vspace{-0.5cm}
Both of these contributions are expected to be small because of 
cancellations between the up- and down-type quark contributions.

{\noindent {\bf Acknowledgements}}

We thank Walter Giele, Eran Yehudai, Bas Tausk and Keith Ellis
for collaboration in the earlier stages of this work.
JMC thanks the UK Particle Physics and Astronomy Research Council
for the award of a research studentship.

\end{document}